\documentclass{article}
\usepackage{amsmath}

\newtheorem{theorem}{Theorem}

\newtheorem{proposition}[theorem]{Proposition}
\newtheorem{remark}[theorem]{Remark}

\newenvironment{proof}[1][Proof]{\textbf{#1.} }{\ \rule{0.5em}{0.5em}}
\input{tcilatex}

\begin{document}

\title{Separable Hamiltonian equations on Riemann manifolds and related
integrable hydrodynamic systems}
\author{Maciej B\l aszak \\
Institute of Physics, A.Mickiewicz University,\\
Umultowska 85, 61-614 Pozna\'{n}, Poland \and Wen-Xiu Ma \\
Department of Mathematics, City University of Hong Kong,\\
Hong Kong, PR China}
\maketitle

\begin{abstract}
A systematic construction of St\"{a}ckel systems in separated coordinates
and its relation to bi-Hamiltonian formalism are considered. A general form
of related hydrodynamic systems, integrable by the Hamilton-Jacobi method,
is derived. One Casimir bi-Hamiltonian case is studed in details and in this
case, a systematic construction of related hydrodynamic systems in arbitrary
coordinates is presented, using the cofactor method and soliton symmetry
constraints.


\end{abstract}

\section{Introduction}

There is a quite well developed theory of the passage from an integrable,
infinite dimensional Hamiltonian system (soliton system) to its various
constrained flows which are themselves completely integrable Hamiltonian
systems. Actually, by using the Hamilton-Jacobi method with respect to two
evolution parameters $x$ and $t$, $N$-gap solutions and $N$-soliton
solutions of a given PDE can be constructed directly from solutions of
related ODEs (constrained flows) \cite{m1}-\cite{m4}.

In the present paper we are interested in, instead of soliton systems, the
first order quasi-linear PDEs \ of the form 
\begin{equation}
q_{it}=\sum_{j=1}^{n}w_{ij}(q)q_{jx},\,\;\;\;\;q_{i}=q_{i}(x,t),\;%
\;i=1,...,n,  \tag{1.1}  \label{1}
\end{equation}%
called \emph{hydrodynamic} or \emph{dispersionless} systems. More precisely,
we consider these systems among (\ref{1}), whose general solutions can be
obtained from the solutions of the related integrable finite dimensional
Hamiltonian systems, like in the case of soliton systems. Such an approach
was presented for the first time by Ferapontov and Fordy in \cite{f1}, where
authors gave general solutions of appropriate hydrodynamic systems from
general solutions of related separable finite dimensional systems.\ The idea
is the following. Consider for example a completely integrable Hamiltonian
system of two degrees of freedom, given by a Hamiltonian function 
\begin{equation}
H=\frac{1}{2}(p_{1}^{2}+p_{2}^{2})+V(q_{1},q_{2})  \tag{1.2}  \label{2}
\end{equation}
and an additional constant of motion 
\begin{equation}
F=q_{2}p_{1}p_{2}-q_{1}p_{2}^{2}+W(q_{1},q_{2}),  \tag{1.3}  \label{3}
\end{equation}%
which commutes with $H$ with respect to the canonical Poisson bracket. Note
that both functions are quadratic in momenta, which belongs to the class to
be considered in this paper. Let $x$ be an evolution parameter of the flow
generated by $H$ and $t$ be an evolution parameter of the flow generated by $%
F$. The commutativity of the two flows means that we can consider a
2-dimensional surface in phase space, parametrized by $x$ and $t.$ On this
surface, the equations of motion for $q$ are 
\begin{equation}
q_{ix}=\frac{\partial H}{\partial p_{i}},\,\;\;q_{it}=\frac{\partial F}{%
\partial p_{i}},\;\;i=1,2.  \tag{1.4}  \label{4}
\end{equation}
Eliminating $p_{1}$ and $p_{2}$ from (\ref{4}), we obtain a system of
hydrodynamic type (\ref{1}) 
\begin{equation}
q_{1t}=q_{2}q_{2x},\;\;\;q_{2t}=q_{2}q_{1x}-2q_{1}q_{2x}.  \tag{1.5}
\label{5}
\end{equation}%
In this calculation, $V$ and $W$ play no role, so in fact the hydrodynamic
system is generated by the geodesic parts of both Hamiltonian functions. The
functions $V,W$ for which $H$ and $F$ commute belong to the St\"{a}ckel
class of parabolic coordinates and are the conserved density and flux for
the hydrodynamic system respectively, as $V_{t}=W_{x}.$ Moreover, as there
is an infinite hierarchy of separable potentials $V,W,$ we have an infinite
hierarchy of conserved densities and related fluxes. Hence, solutions of
Hamiltonian systems (\ref{2}) with arbitrary separable potential $V$ are
simultaneous solutions of hydrodynamic system (\ref{5}).

\begin{remark}
One can pass to higher order PDEs eliminating one of the \ $q_{i}$ but it is
necessary to specify a particular form of $V$. The H\'{e}non-Heiles
potential $V=q_{1}^{3}+\frac{1}{2}q_{1}q_{2}^{2}$ leads to the KdV equation
for $q_{1}$ \cite{f1}, whilst the quartic potential $%
V=16q_{1}^{4}+12q_{1}^{2}q_{2}^{2}+q_{2}^{4}$ generates 
\begin{equation*}
q_{1t}=-\frac{1}{48}\left( \frac{q_{1xx}}{q_{1}}+64q_{1}^{2}\right) _{x},
\end{equation*}
which is not integrable PDE (all the integrable cases of such equations are
listed in canonical form in \cite{mi}).
\end{remark}

Here following this line we consider the problem more systematically. First,
we observe that all examples from \cite{f1} belong to subclass of St\"{a}%
ckel systems considered by Benenti \cite{be}, which, as was shown by Ibort,
Magri and Marmo \cite{mg2}, are one-Casimir bi-Hamiltonian systems. Hence,
applying one-Casimir and developing multi-Casimir bi-Hamiltonian formalism
to quadratic in momenta separable systems, we were able to construct in very
systematic way related hydrodynamic systems together with their general
solutions. Actually, in Section 2 we derive explicitly, in separated
coordinates, the general form of St\"{a}ckel systems on Riemann
(pseudo-Riemann) manifold and the related hydrodynamic systems integrable by
Hamilton-Jacobi method. Then, in Section 3, we connect the considered St\"{a}%
ckel systems with a bi-Hamiltonian formalism generalizing one-Casimir case
onto multi-Casimir one. In Section 4 we present a one-Casimir bi-Hamiltonian
systems in arbitrary coordinates (not necessary canonical), which we call
the cofactor St\"{a}ckel systems, and related cofactor hydrodynamic
counterparts. Then, we present a recipe for the construction of some class
of cofactor hydrodynamic systems in the Cartesian coordinate frame. Finally,
in Section 5, we use constrained flows of soliton systems for a systematic
derivation of hydrodynamic systems from the class considered.

\section{From St\"{a}ckel Hamiltonians to complete integral of related
hydrodynamic systems}

All examples discussed in this paper belong to the class of separable
systems associated with St\"{a}ckel matrices. Actually, in 1893 St\"{a}ckel
gave the first characterization of the Riemann (pseudo-Riemann) manifold $%
(Q,g)$ on which the equations of geodesic motion can be solved by separation
of variables. He proved that if in a system of orthogonal coordinates $%
(\lambda ,\mu )$ there exists a non-singular matrix $\varphi =(\varphi
_{k}^{l}(\lambda _{k}))$, called a \emph{St\"{a}ckel matrix} such that the
geodesic Hamiltonians $E_{r}$ are of the form 
\begin{equation}
E_{r}=\sum_{i=1}^{n}(\varphi ^{-1})_{r}^{i}\mu _{i}^{2},  \tag{2.1}
\label{a}
\end{equation}%
then $E_{r}$ are functionally independent, pairwise commute with respect to
the canonical Poisson bracket and the Hamilton-Jacobi equation associated to 
$E_{1}$ is separable.

Then, Eisenhart gave a coordinate-free representation for St\"{a}ckel
geodesic motion introducing special family of \emph{Killing tensors}. He
proved \cite{Ei} that the geodesic Hamiltonians can be transformed into a St%
\"{a}ckel form (\ref{a}) if the contravariant metric tensor $G=g^{-1}$ has $%
(n-1)$ commuting independent contravariant Killing tensors $A_{r}$ of a
second order such that 
\begin{equation}
E_{r}=\sum_{i,j}A_{r}^{ij}p_{i}p_{j},  \tag{2.2}  \label{b}
\end{equation}%
admitting a common system of closed eigenforms $\alpha _{i}$ 
\begin{equation}
(A_{r}^{\ast }-v_{r}^{i}G)\alpha _{i}=0,\;\;d\alpha _{i}=0,\;\;i=1,...,n, 
\tag{2.3}  \label{c}
\end{equation}%
where $v_{r}^{i}$ are eigenvalues of $(1,1)$ Killing tensor $K_{r}=A_{r}g$ $%
(K_{r}^{\ast }=gA_{r}^{\ast }).$ In local coordinates $q$ on $Q$ we have 
\begin{equation}
K_{r}=\sum_{i,j}(K_{r})_{j}^{i}\frac{\partial }{\partial q_{i}}\otimes
dq_{j},\;\;K_{r}^{\ast }=\sum_{i,j}(K_{r})_{i}^{j}dq_{i}\otimes \frac{%
\partial }{\partial q_{j}}.  \tag{2.4}  \label{aa}
\end{equation}

From now on, separated canonical coordinates will be denoted by $(\lambda
,\mu )$ and natural coordinates, not necessarily canonical, by $(q,p).$ For $%
n$ degrees of freedom, let us consider $n$ St\"{a}ckel Hamiltonian functions
in separated coordinates in the following form 
\begin{equation}
H_{r}=\sum_{i=1}^{n}v_{r}^{i}g^{ii}\mu _{i}^{2}+V_{r}(\lambda )=\mu
^{T}K_{r}G\mu +V_{r}(\lambda ),\;\;\;r=1,...,n,  \tag{2.5}  \label{6}
\end{equation}%
where $\mu =(\mu _{1},...,\mu _{n})^{T},$and $V_{r}(\lambda )$ are
appropriate potentials separable in $(\lambda ,\mu )$ coordinates. So, we
have an $n$-dimensional surface parametrized by $n$ Hamiltonian ''times'' $%
t_{1}=x,t_{2},...,t_{n}.$ In this case we have $\frac{1}{2}n(n-1)$
hydrodynamic systems (\ref{1}) written down in the form 
\begin{equation}
\lambda _{it_{k}}=w_{kl}^{i}\lambda
_{it_{l}},\;\;\;w_{kl}^{i}=v_{k}^{i}/v_{l}^{i},\;\;\;k>l=1,2,...,n-1, 
\tag{2.6}  \label{10}
\end{equation}%
and the corresponding conservation laws: $V_{lt_{k}}=V_{kt_{l}}.$ For a
given evolution parameter $t_{k}$ and a ''space'' coordinate $t_{l},$ there
are $n-2$ hydrodynamic flows $\lambda _{it_{r}}=w_{kl}^{i}\lambda
_{it_{l}},r\neq k\neq l$ which commute with a given flow. It follows
directly from the involutivity of (\ref{6}). The requirement that $H=H_{1},$
and $H_{k}$ are in involution with respect to the canonical Poisson bracket
leads to the equations: 
\begin{equation}
\partial _{i}v_{k}^{i}=0\;\;\;\text{for any }i=1,...,n,  \tag{2.7}  \label{7}
\end{equation}%
\begin{equation}
\partial _{j}\ln (g^{ii})=\frac{\partial _{j}v_{k}^{i}}{v_{k}^{j}-v_{k}^{i}}%
\;\;\text{for any \ }i\neq j,  \tag{2.8}  \label{8}
\end{equation}%
\begin{equation}
\partial _{i}V_{k}-v_{k}^{i}\partial _{i}V_{1}=0\;\;\text{for any \ }%
i=1,...,n,  \tag{2.9}  \label{9}
\end{equation}%
where $\partial _{i}:=\partial /\partial q_{i}.$ Condition (\ref{7}) means
that our system (\ref{10}) is \emph{linearly degenerate}. Cross
differentiation of (\ref{8}) gives 
\begin{equation}
\partial _{l}\left( \frac{\partial _{j}v_{k}^{i}}{v_{k}^{j}-v_{k}^{i}}%
\right) =\partial _{j}\left( \frac{\partial _{l}v_{k}^{i}}{%
v_{k}^{l}-v_{k}^{i}}\right) \;\;\text{for any \ }i\neq j\neq l\neq i. 
\tag{2.10}  \label{11}
\end{equation}%
This last condition is called the ''\emph{semi-Hamiltonian}'' property \cite%
{ts} in the context of systems of hydrodynamic type. In that sense we
consider weakly nonlinear semi-Hamiltonian (WNSH) hydrodynamic systems.

The solution of the system of equations (\ref{7})-(\ref{9}) is easy to
derive if one realizes that the separated coordinates $(\lambda ,\mu )$ for
a Liouville integrable system have to fulfil the Sklyanin conditions \cite%
{sk} 
\begin{equation}
\varphi _{i}(\lambda _{i},\mu _{i};H_{1},...,H_{n})=0,\;\;\;i=1,...,n, 
\tag{2.11}  \label{12}
\end{equation}%
which guarantees the solvability of an appropriate Hamilton-Jacobi equation.
For the integrable system (\ref{6}), under the assumption that all functions 
$\varphi _{i}$\ are linear with respect to all $H_{j},$ conditions (\ref{12}%
) take the general form 
\begin{equation}
f_{i}(\lambda _{i})\mu _{i}^{2}+\gamma _{i}(\lambda _{i})=\sum_{k=1}^{n}\Phi
_{i}^{k}(\lambda _{i})H_{k},\;\;\;i=1,...,n,  \tag{2.12}  \label{13}
\end{equation}%
where $f_{i},\gamma _{i},\Phi _{i}^{k}$ are arbitrary smooth functions of
its argument and the normalization $\Phi _{i}^{n}=1,\;i=1,...,n$ is assumed.
To get the explicit form of $H_{k}=H_{k}(\lambda ,\mu )$ one has to solve
the system of linear equations (\ref{13}). The results are the following 
\begin{equation}
g^{ii}=(-1)^{i+1}\frac{f_{i}(\lambda _{i})\det W^{i1}}{\det W}%
,\;\;\;v_{r}^{i}=(-1)^{r+1}\frac{\det W^{ir}}{\det W^{i1}},  \tag{2.13}
\label{14}
\end{equation}%
\begin{equation}
V_{r}=\sum_{i=1}^{n}(-1)^{i+r}\gamma _{i}(\lambda _{i})\frac{\det W^{ir}}{%
\det W},  \tag{2.14}  \label{rr}
\end{equation}%
where 
\begin{equation}
W=\left( 
\begin{array}{ccccc}
\Phi _{1}^{1}(\lambda _{1}) & \Phi _{1}^{2}(\lambda _{1}) & \cdots & \Phi
_{1}^{n-1}(\lambda _{1}) & 1 \\ 
\vdots & \vdots & \cdots & \vdots & \vdots \\ 
\Phi _{n}^{1}(\lambda _{n}) & \Phi _{n}^{2}(\lambda _{n}) & \cdots & \Phi
_{n}^{n-1}(\lambda _{n}) & 1%
\end{array}%
\right)  \tag{2.15}  \label{u}
\end{equation}%
and $W^{ik}$ is the $(n-1)\times (n-1)$ matrix obtained from $W$ after we
cancel its $i$th row and $k$th column. Then the St\"{a}ckel matrix $\varphi $
is given by 
\begin{equation}
\varphi =\left( 
\begin{array}{ccccc}
\frac{\Phi _{1}^{1}(\lambda _{1})}{f_{1}(\lambda _{1})} & \frac{\Phi
_{1}^{2}(\lambda _{1})}{f_{1}(\lambda _{1})} & \cdots & \frac{\Phi
_{1}^{n-1}(\lambda _{1})}{f_{1}(\lambda _{1})} & \frac{1}{f_{1}(\lambda _{1})%
} \\ 
\vdots & \vdots & \cdots & \vdots & \vdots \\ 
\frac{\Phi _{n}^{1}(\lambda _{n})}{f_{n}(\lambda _{n})} & \frac{\Phi
_{n}^{2}(\lambda _{n})}{f_{n}(\lambda _{n})} & \cdots & \frac{\Phi
_{n}^{n-1}(\lambda _{n})}{f_{n}(\lambda _{n})} & \frac{1}{f_{n}(\lambda _{n})%
}%
\end{array}%
\right) .  \tag{2.16}  \label{d}
\end{equation}%
Notice that for $r=2,$ we reconstructed the result of Ferapontov \cite{f2}
for the functions $v_{2}^{i}$.

\begin{remark}
One can observe that for all known separable systems, we have $f_{i}=f,$ $%
\gamma _{i}=\gamma ,$ $\Phi _{i}^{k}=\Phi ^{k},$ $i=1,...,n,$ so the
conditions (\ref{13}) are represented by the separation (spectral) curve 
\begin{equation*}
f(\lambda )\mu ^{2}+\gamma (\lambda )=\sum_{k=1}^{n}\Phi ^{k}(\lambda )H_{k}.
\end{equation*}
\end{remark}

Given a Hamiltonian system in canonical separated coordinates $(\lambda ,\mu
)$ we can linearize the system through a canonical transformation $(\lambda
,\mu )\rightarrow (b,a)$ in the form $b_{i}=\frac{\partial S}{\partial a_{i}}%
,\mu _{i}=\frac{\partial S}{\partial \lambda _{i}},$ where $S(\lambda
,a)=\sum_{i=1}^{n}S_{i}(\lambda _{i},a)$ is an additively separated
generating function, satisfying the related Hamilton-Jacobi equations 
\begin{equation}
H_{r}(\lambda ,\frac{\partial S}{\partial \lambda })=a_{r}\,,\;\;\;r=1,...,n.
\tag{2.17}  \label{15}
\end{equation}
For the Hamiltonian functions (\ref{6}), fulfilling the Sklyanin conditions (%
\ref{13}), $S_{i}(\lambda _{i},a)$ are given by a system of ordinary
differential equations 
\begin{equation}
f_{i}(\lambda _{i})\left( \frac{dS_{i}}{d\lambda _{i}}\right) ^{2}+\gamma
_{i}(\lambda _{i})=\sum_{k=1}^{n}\Phi _{i}^{k}(\lambda
_{i})a_{k},\;\;\;i=1,...,n.  \tag{2.18}  \label{16}
\end{equation}
Then, in $(b,a)$ coordinates the flows are trivial 
\begin{equation}
(a_{j})_{t_{r}}=0,\;\;(b_{j})_{t_{r}}=\delta _{jr}  \tag{2.19}  \label{17}
\end{equation}
and the implicit form of the trajectories $\lambda _{i}(t_{1},...,t_{n})$ is 
\begin{equation}
b_{j}(\lambda ,a)=\int^{\lambda _{1}}\frac{\Phi _{1}^{n-j}(\xi )}{\varphi
_{1}(\xi )}d\xi +...+\int^{\lambda _{n}}\frac{\Phi _{n}^{n-j}(\xi )}{\varphi
_{n}(\xi )}d\xi =t_{j}+const_{j},\;\;j=1,...,n,\;  \tag{2.20}  \label{18}
\end{equation}
where 
\begin{equation}
\varphi _{i}(\xi )=\left( f_{i}(\xi )\left[ \sum_{k=1}^{n}\Phi _{i}^{k}(\xi
)a_{k}-\gamma _{i}(\xi )\right] \right) ^{1/2}.  \tag{2.21}  \label{19}
\end{equation}
Multicomponent functions $\lambda _{i}(t_{1},...,t_{n})$ are simultaneous
solutions of all dynamics defined by $n$ Hamiltonians (\ref{6}), as well as
general solutions of all hydrodynamic systems (\ref{10}) \cite{f2}.

Of course, we would like to distinguish our separable St\"{a}ckel system (%
\ref{6}) and the related hydrodynamic systems written down in any natural
coordinates $(q,p)$. As we know, for one and a half century this problem has
been unsolved in general. Quite recently, there appeared two strong
formalisms which allow us to construct systematically a transformation to
separated coordinates. One formalism is based on Lax representation of the
system considered \cite{sk} and the other one on a bi-Hamiltonian formalism 
\cite{b1}- \cite{mg2}. In what follows we will use some results of the
second formalism.

\section{Bi-Hamiltonian St\"{a}ckel systems in separable coordinates}

An important fact is that the St\"{a}ckel systems (\ref{6}) are
bi-Hamiltonian on the phase space $M=T^{\ast }Q\times R^{k}$, i.e. the
cotangent bundle spanned by $k$ additional Casimir coordinates, where $1\leq
k\leq n.$ Let us assume that $M$ is equipped with a linear Poisson pencil $%
\Pi _{\lambda }=\Pi _{1}-\lambda \Pi _{0}$ of rank $2n,$ i.e. a pair of
Poisson operators (tensors) $\Pi _{i}:T^{\ast }M\rightarrow TM$ each of rank 
$2n$ such that their linear combination $\Pi _{1}-\lambda \Pi _{0}$ is again
a Poisson operator for any $\lambda \in R$ ( the operators $\Pi _{0}$ and $%
\Pi _{1}$ are then said to be compatible). Moreover, let us assume that $k$
Casimirs $H_{\lambda }^{(i)},i=1,...,k$ of the pencil $\Pi _{\lambda }$ are
polynomials in $\lambda $ of orders $n_{1},...,n_{k}$ 
\begin{equation}
H_{\lambda }^{(i)}=H_{0}^{(i)}\lambda ^{n_{i}}+H_{1}^{(i)}\lambda
^{n_{i}-1}+...+H_{n_{i}}^{(i)},\;\;\;\;i=1,...,k,  \tag{3.1}  \label{20}
\end{equation}%
where $n_{1}+...+n_{k}=n.$ By expanding equations $\Pi _{\lambda
}(dH_{\lambda }^{(i)})=0,i=1,...,k$ in powers of $\lambda $ and comparing
the coefficients of equal powers, we obtain the following bi-Hamiltonian
chains%
\begin{eqnarray}
\Pi _{0}(dH_{0}^{(i)}) &=&0  \notag \\
\Pi _{0}(dH_{1}^{(i)}) &=&\Pi _{1}(dH_{0}^{(i)})  \notag \\
&&\vdots  \TCItag{3.2}  \label{aaa} \\
\Pi _{0}(dH_{n_{i}}^{(i)}) &=&\Pi _{1}(dH_{n_{i}-1}^{(i)})  \notag \\
0 &=&\Pi _{1}(dH_{n_{i}}^{(i)}),  \notag
\end{eqnarray}%
where $i=1,...,k.$ Notice that each chain starts with a Casimir of the first
Poisson operator and terminates with a Casimir of the second Poisson
operator. As follows from (3.2), the functions $H_{j}^{(i)}$ are in
involution with respect to both Poisson structures. If additionally all $%
H_{j}^{(i)}$ are functionally independent, then the chains define a
Liouville integrable system. Let us introduce the following Casimir
coordinates $c_{i}=H_{0}^{(i)},i=1,...,k.$ From the definition, a
Darboux-Nijenhuis (DN) separated coordinates $(\lambda ,\mu ,c)$ are the
canonical coordinates in which both Poisson structures take the following
form%
\begin{eqnarray}
\Pi _{0} &=&\left( 
\begin{array}{ccccc}
0 & I & 0 & \cdots & 0 \\ 
-I & 0 & 0 & \cdots & 0 \\ 
0 & 0 &  &  &  \\ 
\vdots & \vdots &  & 0 &  \\ 
0 & 0 &  &  & 
\end{array}%
\right) ,  \notag \\
&&  \TCItag{3.3}  \label{bbb} \\
\Pi _{1} &=&\left( 
\begin{array}{ccccc}
0 & \Lambda & \partial H_{1}^{(1)}/\partial \mu & \cdots & \partial
H_{1}^{(k)}/\partial \mu \\ 
-\Lambda & 0 & -\partial H_{1}^{(1)}/\partial \lambda & \cdots & -\partial
H_{1}^{(k)}/\partial \lambda \\ 
\ast & \ast &  &  &  \\ 
\vdots & \vdots &  & 0 &  \\ 
\ast & \ast &  &  & 
\end{array}%
\right) ,  \notag
\end{eqnarray}
where $I$ is $n\times n$ unit matrix, $\Lambda =diag(\lambda
_{1},...,\lambda _{n})$ and the symbol $\ast $ denotes the elements that
make the matrix skew-symmetric.

Now, $n$ Sklyanin conditions (\ref{12}), with $n$ functions $%
H_{j}^{(i)},i=1,...,k,$ $j=1,...,n_{i},$ are 
\begin{equation}
f_{i}(\lambda _{i})\mu _{i}^{2}+\overline{\gamma }_{i}(\lambda
_{i})=\sum_{j=1}^{k}\Psi _{i}^{j}(\lambda _{i})H_{\lambda
_{i}}^{(j)},\;\;\;i=1,...,n,  \tag{3.4}  \label{23}
\end{equation}
where $f_{i},\overline{\gamma }_{i},\Psi _{i}^{j}$ are arbitrary smooth
functions of its argument, $H_{\lambda _{i}}^{(j)}$\ are the Casimir
polynomials (\ref{20}) evaluated in $\lambda =\lambda _{i}$ 
\begin{equation}
H_{\lambda _{i}}^{(j)}=c_{j}\lambda _{i}^{n_{j}}+H_{1}^{(j)}\lambda
_{i}^{n_{j}-1}+...+H_{n_{i}}^{(j)},  \tag{3.5}  \label{24}
\end{equation}
and the normalization $\Psi _{i}^{k}=1,\;i=1,...,n$ is assumed. One can
immediately reconstruct conditions (\ref{13}), as $H_{1}=H_{1}^{(1)},$ $%
H_{2}=H_{2}^{(1)},...,H_{n}=H_{n_{k}}^{(k)},$ $\Phi _{i}^{1}(\lambda
_{i})=\lambda _{i}^{n_{1}-1}\Psi _{i}^{1}(\lambda _{i}),$ $\Phi
_{i}^{2}(\lambda _{i})=\lambda _{i}^{n_{1}-2}\Psi _{i}^{1}(\lambda
_{i}),..., $ $\Phi _{i}^{n-1}(\lambda _{i})=\lambda _{i},$ $\gamma
_{i}(\lambda _{i})=\overline{\gamma }_{i}(\lambda
_{i})-\sum_{j=1}^{k}\lambda _{i}^{n_{j}}\Psi _{i}^{j}(\lambda _{i})c_{j}$
and then, from (\ref{14}) we get $H_{j}^{(i)}=H_{j}^{(i)}(\lambda ,\mu ,c).$

The simplest case of one Casimir is determined by the following Sklyanin
conditions: 
\begin{equation}
f_{i}(\lambda _{i})\mu _{i}^{2}+\gamma _{i}(\lambda _{i})=c\lambda
_{i}^{n}+H_{1}\lambda _{i}^{n-1}+...+H_{n},\;\;\;i=1,...,n.  \tag{3.6}
\label{25}
\end{equation}%
Here we put for simplicity $c_{1}=c,H_{k}^{(1)}=H_{k}.$ Notice that related
hydrodynamic systems, written in $\lambda $ coordinates, are completely
described by $v_{r}^{i}$ functions (\ref{14}) which are determined by the
right hand side of the Sklyanin conditions (\ref{13}). Because in one
Casimir case the r.h.s. of eq. (\ref{25}) is fixed, there is a unique set of
functions $v_{k}^{i}.$ Actually, as was found in \cite{b1}, 
\begin{equation}
H_{r}=-\sum_{i=1}^{n}\frac{\partial \rho _{r}}{\partial \lambda _{i}}\frac{%
f_{i}(\lambda _{i})\mu _{i}^{2}+\gamma _{i}(\lambda _{i})}{\Delta _{i}}%
+c\rho _{r}(\lambda ),\;\;\;r=1,...,n,  \tag{3.7}  \label{26}
\end{equation}%
where $\Delta _{i}=\prod_{j\neq i}(\lambda _{i}-\lambda _{j}),$ and $\rho
_{r}(\lambda )$ are coefficients of the characteristic polynomial of $%
\Lambda $ 
\begin{equation}
\det (\lambda I-\Lambda )=(\lambda -\lambda _{1})(\lambda -\lambda
_{2})...(\lambda -\lambda _{n})=\sum_{i=0}^{n}\rho _{i}\lambda ^{i}, 
\tag{3.8}  \label{27}
\end{equation}%
i.e. are Vi\`{e}te polynomials. In the notation of eq.(\ref{6}) it means
that 
\begin{equation}
g^{ii}=\frac{f_{i}(\lambda _{i})}{\Delta _{i}},\;\;\;v_{r}^{i}=-\frac{%
\partial \rho _{r}}{\partial \lambda _{i}},\;\;\;V_{r}=-\sum_{i=1}^{n}\frac{%
\partial \rho _{r}}{\partial \lambda _{i}}\frac{\gamma _{i}(\lambda _{i})}{%
\Delta _{i}},  \tag{3.9}  \label{28}
\end{equation}%
so $K_{r}=diag(v_{r}^{1},...,v_{r}^{n}),\,G=diag(g^{11},...,g^{nn})$ and the
related St\"{a}ckel matrix $\varphi $ takes the form 
\begin{equation}
\varphi =\left( 
\begin{array}{ccccc}
\frac{\lambda _{1}^{n-1}}{f_{1}(\lambda _{1})} & \frac{\lambda _{1}^{n-2}}{%
f_{1}(\lambda _{1})} & \cdots  & \frac{\lambda _{1}}{f_{1}(\lambda _{1})} & 
\frac{1}{f_{1}(\lambda _{1})} \\ 
\vdots  & \vdots  & \cdots  & \vdots  & \vdots  \\ 
\frac{\lambda _{n}^{n-1}}{f_{n}(\lambda _{n})} & \frac{\lambda _{n}^{n-2}}{%
f_{n}(\lambda _{n})} & \cdots  & \frac{\lambda _{n}}{f_{n}(\lambda _{n})} & 
\frac{1}{f_{n}(\lambda _{n})}%
\end{array}%
\right)   \tag{3.10}  \label{e}
\end{equation}%
An additional term $c\rho _{r}(\lambda )$ is related to the new Casimir
coordinate and can be absorbed by the potential. Note that the Killing
tensors $K_{r}$ are given by the cofactor representation 
\begin{equation}
cof(\lambda I-\Lambda )=\sum_{i=0}^{n-1}K_{n-i}\lambda ^{i}.  \tag{3.11}
\label{29}
\end{equation}%
where $cof(A)$ stands for the matrix of cofactors, so that $cof(A)A=(\det
A)I.$ The cofactor nature of $K_{r}$ gives immediately the following
relation 
\begin{equation}
K_{r+1}=\sum_{k=0}^{r}\rho _{k}(\lambda )\Lambda ^{r-k}  \tag{3.12}
\label{f}
\end{equation}%
and vice versa, $K_{r}$ given by the relation (\ref{f}) are of cofactor form
(\ref{29}). The functions $H_{r}$ (\ref{26}) form a single bi-Hamiltonian
chain (3.2), where 
\begin{equation}
\Pi _{0}=\left( 
\begin{array}{ccc}
0 & I & 0 \\ 
-I & 0 & 0 \\ 
0 & 0 & 0%
\end{array}%
\right) ,\;\Pi _{1}=\left( 
\begin{array}{ccc}
0 & \Lambda  & \partial H_{1}/\partial \mu  \\ 
-\Lambda  & 0 & -\partial H_{1}/\partial \lambda  \\ 
\ast  & \ast  & 0%
\end{array}%
\right) .  \tag{3.13}  \label{30}
\end{equation}%
Observe that both Poisson structures can be projected onto $T^{\ast }Q$ 
\begin{equation}
\Theta _{0}=\left( 
\begin{array}{cc}
0 & I \\ 
-I & 0%
\end{array}%
\right) ,\;\Theta _{1}=\left( 
\begin{array}{cc}
0 & \Lambda  \\ 
-\Lambda  & 0%
\end{array}%
\right)   \tag{3.14}  \label{31}
\end{equation}%
and 
\begin{equation}
N=\theta _{1}\theta _{0}^{-1}=\left( 
\begin{array}{cc}
\Lambda  & 0 \\ 
0 & \Lambda 
\end{array}%
\right)   \tag{3.15}  \label{32}
\end{equation}%
is a $(1,1)$ tensor on $T^{\ast }Q$ with a vanishing Nijenhuis torsion. The
operator $N$ is just a lift from $Q$ to $T^{\ast }Q$ of a $(1,1)$ tensor $%
\Lambda $ 
\begin{equation}
\Lambda =\sum_{i}\lambda _{i}\frac{\partial }{\partial \lambda _{i}}\otimes
d\lambda _{i}  \tag{3.16}
\end{equation}%
on $Q$ with a vanishing Nijenhuis torsion \cite{mg2}. Moreover, as 
\begin{equation}
\Lambda ^{\ast }d\lambda _{i}=\lambda _{i}d\lambda _{i},  \tag{3.17}
\end{equation}%
then 
\begin{eqnarray}
K_{r+1}^{\ast }d\lambda _{i} &=&\sum_{k=0}^{r}\rho _{k}(\lambda )(\Lambda
^{\ast })^{r-k}d\lambda _{i}=\sum_{k=0}^{r}\rho _{k}(\lambda )\lambda
_{i}^{r-k}d\lambda _{i}  \notag \\
&=&-\frac{\partial \rho _{r+1}}{\partial \lambda _{i}}d\lambda
_{i}=v_{r+1}^{i}d\lambda _{i},\;  \TCItag{3.18}  \label{bb}
\end{eqnarray}%
and multiplying both sides of eq.(3.18) by $G$ we get 
\begin{equation}
(GK_{r}^{\ast }-v_{r}^{i}G)d\lambda _{i}=0\Leftrightarrow (A_{r}^{\ast
}-v_{r}^{i}G)d\lambda _{i}=0,\;\;\;i=1,...,n,  \tag{3.19}  \label{ee}
\end{equation}%
i.e. the tensorial Eisenhart realization (\ref{c}) of the St\"{a}ckel
results. Moreover, if we define 
\begin{equation}
\overline{\Lambda }:=\frac{1}{2}\mu ^{T}\Lambda G\mu ,  \tag{3.20}
\label{2e}
\end{equation}%
then 
\begin{equation}
\{\overline{\Lambda },E_{1}\}_{\theta _{0}}=aE_{1},\;\;\;\;a=\mu
^{T}G\partial (Tr\,\,\Lambda )/\partial \lambda .  \tag{3.21}  \label{3e}
\end{equation}%
$\Lambda $ of such property is called conformal Killing tensor with the
associated potential in the form of $Tr\,\Lambda .$ Finally, note that $H_{1}
$ and $H_{n}$ are related by 
\begin{equation}
-\rho _{n}\Pi _{0}(dH_{1})=\Pi _{1}(dH_{n}),  \tag{3.22}  \label{33}
\end{equation}%
which is known as the quasi-bi-Hamiltonian representation and is just a
result of the projection of one Casimir Poisson pencil onto a symplectic
leaf of $\Pi _{0}$ \cite{b2}, \cite{mg2}.

Thus, within the class of one Casimir St\"{a}ckel systems, i.e. \emph{%
cofactor St\"{a}ckel systems}, infinitely many systems from that class are
related to $\frac{1}{2}n(n-1)$ hydrodynamic systems (\ref{10}) governed by $%
n $ Killing matrices $K_{r}$\thinspace (\ref{29}) from geodesic
Hamiltonians. For example: $v_{1}^{i}=1,$ $v_{2}^{i}=\lambda
_{i}-\sum_{k=1}^{n}\lambda _{k}$ and $v_{n}^{i}=(-1)^{n}\left(
\prod_{k=1}^{n}\lambda _{k}\right) /\lambda _{i}.$

There exists a sequence of generic separable potentials $V_{r}^{(k)}$ , \ \ $%
k=\pm 1,\pm 2,...,$ which can be added to geodesic Hamiltonians, given by
the following recursion relation \cite{b3} 
\begin{equation}
V_{r}^{(k+1)}=-V_{r+1}^{(k)}+V_{r}^{(1)}V_{1}^{(k)},\;\;\;V_{r}^{(1)}=\rho
_{r},\;\;k=1,2,...,  \tag{3.23}  \label{34}
\end{equation}%
and its inverse 
\begin{equation}
V_{r}^{(-k-1)}=-V_{r-1}^{(-k)}+V_{r}^{(-1)}V_{n}^{(-k)},\;\;\;V_{r}^{(-1)}=%
\rho _{r-1}/\rho _{n},\;\;\;k=1,2,...\;.  \tag{3.24}  \label{35}
\end{equation}%
Using the notation $V_{\lambda }=\sum_{j=0}^{n-1}V_{n-j}\lambda ^{j},$ the
recursion formulas (\ref{34}) and (\ref{35}) can be written in a compact
form 
\begin{equation}
V_{\lambda }^{(k+1)}=\det (\lambda I-\Lambda )V_{1}^{(k)}-\lambda V_{\lambda
}^{(k)}  \tag{3.25}  \label{36}
\end{equation}%
and 
\begin{equation}
V_{\lambda }^{(-k-1)}=\frac{1}{\lambda }\left( \frac{\det (\lambda I-\Lambda
)}{\det \Lambda }V_{n}^{(-k)}-V_{\lambda }^{(-k)}\right) .  \tag{3.26}
\label{37}
\end{equation}%
Potentials $V^{(k)}$ (\ref{34}) and $V^{(-k)}$ (\ref{35}) are generated by
the corresponding monomials $\gamma _{i}(\lambda _{i})=\lambda _{i}^{n+k-1}$
and $\gamma _{i}(\lambda _{i})=\lambda _{i}^{-k}$ from (\ref{25}). The
infinite hierarchy of conservation laws takes the form $\left(
V_{r}^{(k)}\right) _{t_{s}}=\left( V_{s}^{(k)}\right) _{t_{r}},k=\pm 1,\pm
2,...$ .

Now, the complete integral (\ref{18}) of hydrodynamic systems (\ref{10})
with $v_{r}^{i}=-\frac{\partial \rho _{r}}{\partial \lambda _{i}}$ is given
by 
\begin{equation}
\int^{\lambda _{1}}\frac{\xi ^{n-j}}{\varphi _{1}(\xi )}d\xi
+...+\int^{\lambda _{n}}\frac{\xi _{n}^{n-j}}{\varphi _{n}(\xi )}d\xi
=t_{j}+const_{j},\;\;j=1,...,n,  \tag{3.27}  \label{38}
\end{equation}%
where the functions $\varphi _{i}(\xi )$ are arbitrary. Additionally, one
can always construct a $(1+(n-1))$-dimensional hydrodynamic system with the
solution of the form (\ref{38}) involving all independent variables $%
t_{j},j=1,...,n$ simultaneously.

\textbf{Example 1 }Three-field system $(\lambda _{1},\lambda _{2},\lambda
_{3}).$

In this case, we have 
\begin{eqnarray}
\rho _{1} &=&-\lambda _{1}-\lambda _{2}-\lambda _{3},  \notag \\
\rho _{2} &=&\lambda _{1}\lambda _{2}+\lambda _{1}\lambda _{3}+\lambda
_{2}\lambda _{3},  \TCItag{3.28}  \label{39} \\
\rho _{3} &=&-\lambda _{1}\lambda _{2}\lambda _{3},  \notag
\end{eqnarray}%
hence the following three WNSH hydrodynamic systems in Riemann invariant
form are admissible: 
\begin{equation}
\lambda _{1t_{2}}=-(\lambda _{2}+\lambda _{3})\lambda _{1t_{1}},\;\lambda
_{2t_{2}}=-(\lambda _{1}+\lambda _{3})\lambda _{2t_{1}},\;\lambda
_{3t_{2}}=-(\lambda _{1}+\lambda _{2})\lambda _{3t_{1}},  \tag{3.29}
\label{40}
\end{equation}%
\begin{equation}
\lambda _{1t_{3}}=\lambda _{2}\lambda _{3}\lambda _{1t_{1}},\;\;\lambda
_{2t_{3}}=\lambda _{1}\lambda _{3}\lambda _{2t_{1}},\;\;\lambda
_{3t_{3}}=\lambda _{1}\lambda _{2}\lambda _{3t_{1}},\;\;  \tag{3.30}
\label{41}
\end{equation}%
\begin{equation}
\lambda _{1t_{3}}=-\frac{\lambda _{2}\lambda _{3}}{\lambda _{2}+\lambda _{3}}%
\lambda _{1t_{2}},\;\lambda _{2t_{3}}=-\frac{\lambda _{1}\lambda _{3}}{%
\lambda _{1}+\lambda _{3}}\lambda _{2t_{2}},\;\lambda _{3t_{3}}=-\frac{%
\lambda _{1}\lambda _{2}}{\lambda _{1}+\lambda _{2}}\lambda _{3t_{2}}. 
\tag{3.31}  \label{42}
\end{equation}%
The complete integral for all these systems is given by 
\begin{equation}
\int^{\lambda _{1}}\frac{\xi ^{3-j}}{\varphi _{1}(\xi )}d\xi +\int^{\lambda
_{2}}\frac{\xi ^{3-j}}{\varphi _{2}(\xi )}d\xi +\int^{\lambda _{3}}\frac{\xi
^{3-j}}{\varphi _{3}(\xi )}d\xi =t_{j}+const_{j},\;\;j=1,2,3.  \tag{3.32}
\label{43}
\end{equation}%
In each of the above cases, only a pair of $t_{i}$ coordinates is involved,
so the third one can be put equal to zero. Notice that the first two
equations can be coupled into a $(1+2)$-dimensional hydrodynamic system 
\begin{eqnarray*}
\lambda _{1t_{3}} &=&-\frac{1}{2}(\lambda _{2}^{2}+\lambda _{3}^{2})\lambda
_{1t_{1}}-\frac{1}{2}(\lambda _{2}+\lambda _{3})\lambda _{1t_{2}}, \\
\lambda _{2t_{3}} &=&-\frac{1}{2}(\lambda _{1}^{2}+\lambda _{3}^{2})\lambda
_{2t_{1}}-\frac{1}{2}(\lambda _{1}+\lambda _{3})\lambda _{2t_{2}}, \\
\lambda _{3t_{3}} &=&-\frac{1}{2}(\lambda _{1}^{2}+\lambda _{2}^{2})\lambda
_{3t_{1}}-\frac{1}{2}(\lambda _{1}+\lambda _{2})\lambda _{3t_{2}},
\end{eqnarray*}%
for which (\ref{43}) is an integral involving simultaneously a triple of
independent coordinates $t_{i},i=1,2,3.$

The two-Casimir case is given by the following Sklyanin conditions 
\begin{eqnarray}
f_{i}(\lambda _{i})\mu _{i}^{2}+\gamma _{i}(\lambda _{i}) &=&\Psi
_{i}^{1}(c_{1}\lambda _{i}^{n_{1}}+H_{1}^{(1)}\lambda
_{i}^{n_{1}-1}+...+H_{n_{1}}^{(1)})  \TCItag{3.33}  \label{44} \\
&&+c_{2}\lambda _{i}^{n_{2}}+H_{1}^{(2)}\lambda
_{i}^{n_{2}-1}+...+H_{n_{2}}^{(2)},  \notag
\end{eqnarray}%
where $i=1,...,n$ and $n_{1}+n_{2}=n.$ Because $v_{r}^{i}$ are determined by
the r.h.s. of (3.31), involving arbitrary functions $\Psi _{i}^{1},$ we have
infinitely many sets of $v_{r}^{i}$ functions and infinitely many different
WNSH hydrodynamic systems written in $\lambda $ coordinates, integrable by
the Hamilton-Jacobi method. A particular example, with $n_{1}=1,n_{2}=n-1$
and $\Psi _{i}^{1}=\lambda _{i}^{n},$ can be found in \cite{b2}.

As mentioned before, we are interested in constructing hydrodynamic systems
and a hierarchy of conservation laws, written down in some natural
coordinates, which are integrable by the Hamilton-Jacobi method. Then, we
would like to find an appropriate transformation to Riemann invariant form (%
\ref{10}) represented by separated coordinates. In the next section, we
present some \ \ results, based on the known theory on the separable one
Casimir bi-Hamiltonian chains \cite{l1}-\cite{bm}.

\section{Separable cofactor systems in arbitrary coordinates}

As most relations presented in the previous section, although derived in
separated coordinates, are of tensorial form, i.e. coordinate-free form so
they are valid in arbitrary coordinate frame spanning $Q.$ First, let us
restrict our considerations to a class of one-Casimir bi-Hamiltonian systems
written in arbitrary canonical coordinates governed by a nondegenerate point
transformation between $\lambda $ and $q$ coordinates. After the
transformation, one gets the following Poisson structures 
\begin{equation}
\Pi _{0}=\left( 
\begin{array}{ccc}
0 & I & 0 \\ 
-I & 0 & 0 \\ 
0 & 0 & 0%
\end{array}%
\right) ,\;\Pi _{1}=\left( 
\begin{array}{ccc}
0 & L(q) & \partial H_{1}/\partial p \\ 
-L^{T}(q) & F(q,p) & -\partial H_{1}/\partial q \\ 
\ast & \ast & 0%
\end{array}%
\right) ,  \tag{4.1}  \label{45}
\end{equation}%
where $F_{ij}=\frac{\partial }{\partial q_{i}}(Lp)_{j}-\frac{\partial }{%
\partial q_{j}}(p^{T}L)_{i}$ and a Casimir of the pencil $\Pi _{\lambda
}=\Pi _{1}-\lambda \Pi _{0}$ of the following form 
\begin{eqnarray}
H_{\lambda }(q,p,c) &=&\sum_{i=0}^{n}H_{n-i}(q,p,c)\lambda ^{i}  \TCItag{4.2}
\label{46} \\
&=&p^{T}cof(I\lambda -L)G(q)p+V_{\lambda }^{(\pm k)}(q)+c\det (\lambda I-L),
\notag
\end{eqnarray}%
where separable potentials are generated by the recursions 
\begin{equation}
V_{\lambda }^{(k+1)}(q)=\det (\lambda I-L)V_{1}^{(k)}(q)-\lambda V_{\lambda
}^{(k)}(q),  \tag{4.3}  \label{47}
\end{equation}%
\begin{equation}
V_{\lambda }^{(-k-1)}(q)=\frac{1}{\lambda }\left( \frac{\det (\lambda I-L)}{%
\det L}V_{n}^{(-k)}(q)-V_{\lambda }^{(-k)}(q)\right) .  \tag{4.4}  \label{48}
\end{equation}%
The geodesic Hamiltonians are as follows 
\begin{equation}
E_{r}(q,p)=p^{T}A_{r}(q)p=p^{T}K_{r}(q)G(q)p,\;\;\;%
\sum_{i=0}^{n-1}K_{n-i}(q)\lambda ^{i}=cof(\lambda I-L(q)),  \tag{4.5}
\label{49}
\end{equation}%
with a contravariant metric tensor $G(q)$ and 
\begin{equation}
K_{r}=\sum_{k=0}^{r}\rho _{k}(q)L^{r-k},  \tag{4.6}
\end{equation}%
where $\rho _{r}(q)$ are coefficients of a characteristic polynomial of $%
L(q) $. Note, that cofactor St\"{a}ckel systems are exactly those considered
by Benenti \cite{be},\cite{be1},\cite{mg2}.

Then, the hydrodynamic equations in $q$ representation take the form 
\begin{equation}
q_{t_{j}}=K_{j}K_{i}^{-1}q_{t_{i}},\;\;\;j>i=1,...,n-1.  \tag{4.7}
\label{50}
\end{equation}%
This class of hydrodynamic systems will be called \emph{cofactor
hydrodynamic systems}. Observe, that to get cofactor hydrodynamic systems (%
\ref{50}), we only need a $(1,1)$ conformal Killing tensor $L(q)$. Moreover,
the separated coordinates $\lambda _{i},i=1,...,n$, in which the system (\ref%
{50}) takes the WNSH form (\ref{10}), are given by the roots of 
\begin{equation}
\det (\lambda I-L)=0.  \tag{4.8}  \label{51}
\end{equation}%
Unfortunately, there is no systematic method of constructing $L$ tensors in
a general case. Nevertheless, recently some progress has been made in a
special case of flat spaces and Cartesian $q$ \ coordinates.

Let us consider a class of one-Casimir bi-Hamiltonian systems on a flat
space $Q=R^{n}$, introduced recently in \cite{l1}-\cite{l3}. Here we briefly
review the results which are important for our construction. Let $%
q=(q_{1},...,q_{n})^{T}$ be a set of Cartesian coordinates and $A,$ an $%
n\times n$ matrix, whose elements fulfil the following equations: 
\begin{equation}
\partial _{i}A_{jk}+\partial _{g}A_{ki}+\partial
_{k}A_{ij}=0,\;\;\;i,j,k=1,...,n.  \tag{4.9}  \label{52}
\end{equation}%
The equations (\ref{52}) imply that the matrix $A$ is a Killing matrix. An
important class of solutions of these equations have the form \cite{l3} 
\begin{equation}
A=cof(G)  \tag{4.10}  \label{53}
\end{equation}%
with 
\begin{equation}
G=\alpha qq^{T}+\beta q^{T}+q\beta ^{T}+\gamma ,  \tag{4.11}  \label{54}
\end{equation}%
where $\alpha $ is a real constant, $\beta =(\beta _{1},...,\beta _{n})^{T}$
is a column vector of constants and where $\gamma $ is a symmetric $n\times
n $ matrix. One can show that for $n=2$ it is the general solution of (\ref%
{52}) \cite{f2}.

Now, let 
\begin{equation}
\widetilde{G}=\widetilde{\alpha }qq^{T}+\widetilde{\beta }q^{T}+q\widetilde{%
\beta }^{T}+\widetilde{\gamma }  \tag{4.12}  \label{55}
\end{equation}%
be another matrix of the form (\ref{54}) (we assume that at least one of the
matrices $G$ and $\widetilde{G}$ is nonconstant), then, all matrices $%
A_{i},i=1,...,n,$ defined as coefficients in the polynomial expansion of $%
cof(\widetilde{G}+\lambda G)$ with respect to the real parameter $\lambda :$ 
\begin{equation}
cof(\widetilde{G}+\lambda G)=\sum_{i=0}^{n-1}A_{n-i}\lambda ^{i}  \tag{4.13}
\label{56}
\end{equation}%
with $A_{1}=cof(G),\,A_{n}=cof(\widetilde{G}),$ are Killing matrices.

Let us assume that over some region of $Q$ $\det G\neq 0.$ Then $L=-%
\widetilde{G}G^{-1},$ considered as $(1,1)$ tensor field on $Q=R^{n}$, has a
vanishing Nijenhuis torsion, and so it can always be diagonalized. Moreover,
on $T^{\ast }Q$ we define $n$ functions 
\begin{equation}
E_{i}=p^{T}A_{i}p,\;i=1,...,n  \tag{4.14}  \label{57}
\end{equation}
and a tensor 
\begin{equation}
\Theta =\left( 
\begin{array}{cc}
0 & -G \\ 
G & F%
\end{array}
\right) ,  \tag{4.15}  \label{58}
\end{equation}
where $(p_{1},...,p_{n})^{T}$ are momenta coordinates and an $n\times n$
matrix $F$ is defined by 
\begin{equation}
F=Np^{T}-pN^{T},\;\;\;N=\alpha q+\beta .  \tag{4.16}  \label{59}
\end{equation}

\begin{theorem}
Assuming that over some region of $Q$ the operator $L$ has n functionally
independent nonzero eigenvalues: \ \ \ \ \ \ \ \ \ \ \ \ \ \ \ \ \ \ \ \ \ \
\ \ \ \ \ \ \ \ \ \ \ \ \ 
\end{theorem}

\begin{enumerate}
\item[(i)] $\Theta $ \emph{is a Poisson tensor of rank} $2n.$

\item[(ii)] $E_{i}$ \emph{are functionally independent and in involution
with respect to} $\Theta .$
\end{enumerate}

\begin{proof}
Operator $\Theta $ is skew-symmetric and the Jacobi identity can be proved
by inspection. Moreover $\det G\neq 0$ guarantee its maximal rank. On the
other hand, the cofactor form of geodesic Hamiltonians together with
functional independence of eigenvalues of $L$ operator means that we have
the Eisenhart representation and hence the St\"{a}ckel ones in separated
coordinates.
\end{proof}

Obviously, we have Liouville integrable systems for geodetic motions,
written in noncanonical coordinates $(q,p).$ Choosing $G=-I$ we get a
special case of canonical representations. Before we separate the system,
let us construct related hydrodynamic systems. Equations of motion for $q$
coordinates are as follows 
\begin{equation}
q_{t_{i}}=-2GA_{i}p,\;\;i=1,...,n,  \tag{4.17}  \label{60}
\end{equation}%
hence, elimination of $p$ coordinates, together with the relation $%
A_{1}=cof(G),$ leads to the following cofactor hydrodynamic systems 
\begin{equation}
q_{t_{j}}=A_{1}^{-1}A_{j}A_{i}^{-1}A_{1}q_{t_{i}}=GA_{j}A_{i}^{-1}G^{-1}q_{t_{i}},\;\;j>i=1,...,n-1.
\tag{4.18}  \label{61}
\end{equation}%
In particular case $t_{i}=t_{1}=x$ we have 
\begin{equation}
q_{t_{j}}=A_{1}^{-1}A_{j}q_{x}=\frac{1}{\det G}GA_{j}q_{x},\;\;j=2,...,n. 
\tag{4.19}  \label{62}
\end{equation}

To find separated coordinates for the geodesic Hamiltonians (\ref{57}), we
have to put them into a bi-Hamiltonian form. It can be done on the extended
phase space $M=T^{\ast }Q\times R$ with local coordinates $(q,p,c).$ Let us
introduce functions $D_{j}(q)$ as coefficients in the polynomial expansion
of $\det (\widetilde{G}+\lambda G)$%
\begin{equation}
\sum_{i=0}^{n}D_{n-i}(q)\lambda ^{i}=\det (\widetilde{G}+\lambda G) 
\tag{4.20}  \label{63}
\end{equation}
so that $D_{0}=\det G$ and $D_{n}=\det \widetilde{G}.$

\begin{theorem}
Functions 
\begin{equation}
H_{r}=E_{r}+c\frac{D_{r}}{D_{0}},\;\;r=0,...,n,\;\;E_{0}=c,  \tag{4.21}
\label{64}
\end{equation}
constitute a bi-Hamiltonian chain with respect to a pair of compatible
Poisson structures 
\begin{eqnarray}
\Pi _{0} &=&\left( 
\begin{array}{ccc}
0 & -G & 0 \\ 
G & -F & 0 \\ 
0 & 0 & 0%
\end{array}
\right)  \notag \\
&&  \TCItag{4.22}  \label{65} \\
\Pi _{1} &=&\left( 
\begin{array}{ccc}
0 & \widetilde{G} & -2\det (G)p \\ 
-\widetilde{G} & \widetilde{F} & 2(\widetilde{N}+LN)c \\ 
\ast & \ast & 0%
\end{array}
\right) ,  \notag
\end{eqnarray}
which starts with the Casimir $H_{0}=c$ of the first Poisson structure $\Pi
_{0}$ and terminates with the Casimir $H_{n}=E_{n}+c\frac{D_{n}}{D_{0}}$\ of
the second Poisson structure $\Pi _{1}.$ The last column of $\Pi _{1}$ is a
first vector field from the hierarchy: $\Pi _{0}(dH_{1}).$
\end{theorem}

Having a bi-Hamiltonian chain, one can systematically construct separated
coordinates \cite{bm}. The result is given as follows:

\begin{theorem}
For a geodesic Hamiltonian system (\ref{57}), (\ref{58}) the separated
coordinates $\lambda _{i}(q)$ are the roots of the equation 
\begin{equation}
\det (\lambda I-L)=0\Longleftrightarrow \det (\lambda G+\widetilde{G})=0 
\tag{4.23}  \label{66}
\end{equation}%
and related momenta $\mu _{i}(q,p)$ are given by the equations 
\begin{equation}
\mu _{i}(q,p)=\frac{1}{2}\frac{\Omega ^{T}cof(\widetilde{G}+\lambda
_{i}(q)G)p}{\Omega ^{T}cof(\widetilde{G}+\lambda _{i}(q)G)\Omega }%
,\;\;i=1,...,n  \tag{4.24}  \label{67}
\end{equation}%
where $\Omega =(\widetilde{N}+LN)$ and $\widetilde{N}=\widetilde{\alpha }q+%
\widetilde{\beta }.$ The Poisson operators (4.22) attain the form 
\begin{eqnarray}
\Pi _{0} &=&\left( 
\begin{array}{ccc}
0 & I & 0 \\ 
-I & 0 & 0 \\ 
0 & 0 & 0%
\end{array}%
\right) ,  \notag \\
&&  \TCItag{4.25}  \label{68} \\
\Pi _{1} &=&\left( 
\begin{array}{ccc}
0 & \Lambda & \partial H_{1}/\partial \mu \\ 
-\Lambda & 0 & -\partial H_{1}/\partial \lambda \\ 
\ast & \ast & 0%
\end{array}%
\right)  \notag
\end{eqnarray}%
while the Hamiltonians (\ref{64}) have the form 
\begin{equation}
H_{r}=-\sum_{k=1}^{n}\frac{\partial \rho _{r}}{\partial \lambda _{k}}\frac{%
f_{k}(\lambda _{k})}{\Delta _{k}}\mu _{k}^{2}+c\rho _{r}(\lambda
),\;\;r=1,...,n.  \tag{4.30}  \label{69}
\end{equation}
\end{theorem}

The above theorem ensures us that under the transformation given by the
roots of (\ref{66}), cofactor hydrodynamic systems (\ref{61}) turn into the
WNSH form (\ref{10}), where $(K_{r})_{j}^{i}=\delta _{ij}v_{r}^{j}=-\frac{%
\partial \rho _{r}}{\partial \lambda _{i}}.$ To complete the picture from
the point of view of St\"{a}ckel systems, let us end this consideration with
the following result.

\begin{proposition}
The generic separable potentials of the cofactor St\"{a}ckel system (\ref{64}%
),(4.22) are given by the recursion formulas \cite{l3} 
\begin{eqnarray}
V_{\lambda }^{(k+1)}(q) &=&\frac{\det (\lambda G+\widetilde{G})}{\det G}%
V_{1}^{(k)}(q)-\lambda V_{\lambda }^{(k)}(q)  \TCItag{4.31}  \label{70} \\
&=&\det (\lambda I-L)V_{1}^{(k)}(q)-\lambda V_{\lambda }^{(k)}(q),  \notag
\end{eqnarray}%
\begin{eqnarray}
V_{\lambda }^{(-k-1)}(q) &=&\frac{1}{\lambda }\left( \frac{\det (\lambda G+%
\widetilde{G})}{\det \widetilde{G}}V_{n}^{(-k)}(q)-V_{\lambda
}^{(-k)}(q)\right)  \TCItag{4.32}  \label{71} \\
&=&\frac{1}{\lambda }\left( \frac{\det (\lambda I-L)}{\det L}%
V_{n}^{(-k)}-V_{\lambda }^{(-k)}\right)  \notag
\end{eqnarray}%
where $V_{\lambda }=\sum_{i=0}^{n-1}V_{n-i}\lambda ^{i}$ and the potential $%
V_{r}$ belongs to the appropriate Hamiltonian $E_{r}$ (\ref{64}).
\end{proposition}

In the canonical coordinates $G=-I$, $\widetilde{G}=L$ and we reconstruct
the potentials (\ref{47}) and (\ref{48}).

\textbf{Example 2 }A two field cofactor hydrodynamic system is defined by 
\begin{equation*}
G=\left( 
\begin{array}{cc}
1 & q_{1} \\ 
q_{1} & 2q_{2}%
\end{array}%
\right) ,\;\widetilde{G}=\left( 
\begin{array}{cc}
q_{1}^{2}+1 & q_{1}q_{2} \\ 
q_{1}q_{2} & q_{2}^{2}%
\end{array}%
\right) .
\end{equation*}%
We have 
\begin{equation*}
A_{1}=\left( 
\begin{array}{cc}
2q_{2} & -q_{1} \\ 
-q_{1} & 1%
\end{array}%
\right) ,\;A_{2}=\left( 
\begin{array}{cc}
q_{2}^{2} & -q_{1}q_{2} \\ 
-q_{1}q_{2} & q_{1}^{2}+1%
\end{array}%
\right)
\end{equation*}%
and hence, form (\ref{62}) 
\begin{eqnarray*}
q_{1t_{2}} &=&\frac{q_{2}(q_{2}-q_{1}^{2})}{2q_{2}-q_{1}^{2}}q_{1t_{1}}-%
\frac{q_{1}(q_{2}-q_{1}^{2}-1)}{2q_{2}-q_{1}^{2}}q_{1t_{1}}, \\
q_{2t_{2}} &=&-\frac{q_{2}(q_{1}^{2}+2)}{2q_{2}-q_{1}^{2}}q_{1t_{1}}+\frac{%
q_{2}(q_{1}^{2}+2)}{2q_{2}-q_{1}^{2}}q_{1t_{1}}.
\end{eqnarray*}%
The transformation (\ref{66}) to the WNSH form 
\begin{equation*}
\lambda _{1t_{2}}=\lambda _{2}\lambda _{1t_{1}},\;\;\lambda
_{2t_{2}}=\lambda _{1}\lambda _{2t_{1}},
\end{equation*}%
is given by 
\begin{equation*}
q_{1}=-2\frac{\sqrt{-\lambda _{1}\lambda _{2}(\lambda _{1}+1)(\lambda _{2}+2)%
}}{\lambda _{1}+\lambda _{2}+\lambda _{1}\lambda _{2}},\;\;q_{2}=-2\frac{%
\lambda _{1}\lambda _{2}}{\lambda _{1}+\lambda _{2}+\lambda _{1}\lambda _{2}}%
.
\end{equation*}

\textbf{Example 3 }Three field cofactor hydrodynamic systems are defined by 
\begin{equation*}
G=\left( 
\begin{array}{ccc}
0 & 0 & -1 \\ 
0 & -1 & 0 \\ 
-1 & 0 & 0%
\end{array}%
\right) ,\;\widetilde{G}=\left( 
\begin{array}{ccc}
2q_{3} & q_{2} & q_{1} \\ 
q_{2} & 0 & -1 \\ 
q_{1} & -1 & 0%
\end{array}%
\right) .
\end{equation*}%
We have 
\begin{equation*}
A_{1}=\left( 
\begin{array}{ccc}
0 & 0 & -1 \\ 
0 & -1 & 0 \\ 
-1 & 0 & 0%
\end{array}%
\right) ,\;A_{2}=\left( 
\begin{array}{ccc}
0 & 1 & q_{1} \\ 
1 & 2q_{1} & -q_{2} \\ 
q_{1} & -q_{2} & -2q_{3}%
\end{array}%
\right) ,
\end{equation*}%
\begin{equation*}
A_{2}=\left( 
\begin{array}{ccc}
-1 & -q_{1} & -q_{2} \\ 
-q_{1} & -q_{1}^{2} & 2q_{3}+q_{1}q_{2} \\ 
-q_{2} & 2q_{3}+q_{1}q_{2} & -q_{2}^{2}%
\end{array}%
\right)
\end{equation*}%
and hence, the following hydrodynamic systems 
\begin{eqnarray*}
q_{1t_{2}} &=&-q_{1}q_{1t_{1}}+q_{2}q_{2t_{1}}+2q_{3}q_{3t_{1}}, \\
q_{1t_{2}} &=&-q_{1t_{1}}-2q_{1}q_{2t_{1}}+q_{2}q_{3t_{1}}, \\
q_{1t_{2}} &=&-q_{2}q_{2t_{1}}-q_{1}q_{3t_{1}},
\end{eqnarray*}%
\begin{eqnarray*}
q_{1t_{3}}
&=&q_{2}q_{1t_{1}}-(q_{1}q_{2}+2q_{3})q_{2t_{1}}+q_{2}^{2}q_{3t_{1}}, \\
q_{1t_{3}}
&=&q_{1}q_{1t_{1}}+q_{1}^{2}q_{2t_{1}}-(q_{1}q_{2}+2q_{3})q_{3t_{1}}, \\
q_{1t_{3}} &=&q_{1t_{1}}-q_{1}q_{2t_{1}}+q_{2}q_{3t_{1}},
\end{eqnarray*}%
\begin{eqnarray*}
q_{1t_{3}} &=&\frac{1}{2}\frac{2q_{2}^{2}+3q_{1}^{2}q_{2}+2q_{1}q_{3}}{%
q_{3}-q_{1}q_{2}-q_{1}^{3}}q_{1t_{2}}-\frac{1}{2}\frac{%
2q_{2}q_{3}+2q_{1}^{2}q_{3}+q_{1}^{3}q_{2}}{q_{3}-q_{1}q_{2}-q_{1}^{3}}%
q_{2t_{2}} \\
&&+\frac{1}{2}\frac{4q_{1}q_{2}q_{3}+2q_{2}^{3}+q_{3}^{2}+q_{1}^{2}q_{2}^{2}%
}{q_{3}-q_{1}q_{2}-q_{1}^{3}}q_{3t_{2}}, \\
&&...\;\;.
\end{eqnarray*}%
The transformation (\ref{66}) to the WNSH form (\ref{40})-(\ref{42}) is
given by 
\begin{eqnarray*}
q_{1} &=&-\frac{1}{2}\rho _{1}, \\
q_{2} &=&\frac{1}{2}\rho _{2}-\frac{1}{8}\rho _{1}^{2}, \\
q_{3} &=&-\frac{1}{2}\rho _{3}+\frac{1}{4}\rho _{1}\rho _{2}-\frac{1}{16}%
\rho _{1}^{3},
\end{eqnarray*}%
where $\rho _{i}$ are define by (3.26).

\section{Hydrodynamic equations related to constrained flows of soliton
systems}

As it was mentioned in the introduction, symmetry constraints of soliton
systems give us a systematic method of constructing Liouville integrable
finite dimensional Hamiltonian systems \cite{ma}. Moreover, most of the
examples constructed so far, belongs to separable systems, separated either
through a bi-Hamiltonian formalism \cite{b1}, \cite{m3} or through a
spectral curve method \cite{m1}, \cite{m2}. If additionally involutive
constants of motion are quadratic forms of momenta, then we can again
systematically construct related hydrodynamic systems which have a complete
integral in the form (\ref{38}).

We illustrate the approach by two examples. The first one is related to the
constrained Schr\"{o}dinger spectral problem of the KdV hierarchy, where the
natural coordinates are its $n$ eigenfunctions. Let $q_{k}$ be an
eigenfunction of the KdV Lax operator $\partial _{x}^{2}+u$ with an
eigenvalue $\alpha _{k}$ 
\begin{equation}
q_{kxx}+uq_{k}=\alpha _{k}q_{k},\;\;k=1,...,n.  \tag{5.1}  \label{72}
\end{equation}%
Under the symmetry constraint 
\begin{equation}
u_{x}=\sum_{i=1}^{n}\left( q_{i}^{2}\right) _{x}\Rightarrow
u=\sum_{i=1}^{n}q_{i}^{2}+c  \tag{5.2}  \label{73}
\end{equation}%
of the KdV equation, where $c\,$ plays a role of the additional Casimir
coordinate, we obtain the Garnier system, well known in the classical
mechanics: 
\begin{equation}
q_{kxx}+q_{k}\sum_{i=1}^{n}q_{i}^{2}+cq_{k}=\alpha _{k}q_{k},\;\;k=1,...,n. 
\tag{5.3}  \label{74}
\end{equation}%
The bi-Hamiltonian representation of (\ref{74}), in canonical coordinates,
was found in \cite{st}, \cite{b4}, where the second Poisson structure is 
\begin{equation}
\left( 
\begin{array}{ccc}
0 & A-\frac{1}{2}qq^{T} & p \\ 
\frac{1}{2}qq^{T}-A & \frac{1}{2}pq^{T}-\frac{1}{2}qp^{T} & [A-cI-(q,q)]q \\ 
\ast & \ast & 0%
\end{array}%
\right) ,  \tag{5.4}  \label{75}
\end{equation}%
$p_{k}=q_{kx},k=1,...,n,$ $A=diag(\alpha _{1},...,\alpha _{n})$ and $%
(q,q)=q^{T}q=\sum_{i=1}^{n}q_{i}^{2}.$ Observe that this is again a cofactor
St\"{a}ckel type system where $G=-I$ and $\widetilde{G}=L=A-\frac{1}{2}%
qq^{T}.$ Geodesic motion is separable in generalized elliptic coordinates $%
\lambda _{1},...,\lambda _{n},$ defined by the relation 
\begin{equation}
1+\frac{1}{2}\sum_{k=1}^{n}\frac{q_{k}^{2}}{z-\alpha _{k}}=\frac{%
\prod_{j=1}^{n}(z-\lambda _{j})}{\prod_{j=1}^{n}(z-\alpha _{j})},  \tag{5.5}
\label{76}
\end{equation}%
which are just the DN coordinates defined by the roots of $\det (\lambda G+%
\widetilde{G})=0$ \cite{b1}. The Garnier potential is the first nontrivial
one from generic potentials generated by the recursion (\ref{47}), all are
separable in elliptic coordinates (\ref{76}). The geodesic Hamiltonians are
of the form 
\begin{equation*}
H_{k+1}=\frac{1}{2}(p,A^{k}p)+\frac{1}{4}%
\sum_{j=1}^{k}[(q,A^{j-1}q)(p,A^{k-j}p)-(q,A^{j-1}p)(q,A^{k-j}p)],
\end{equation*}%
where $k=0,...,n-1$ and they allow us to construct related hydrodynamic
systems. For $n=3$ the three cofactor hydrodynamic systems are 
\begin{equation}
q_{t_{2}}=A_{2}q_{x},\;q_{t_{3}}=A_{3}q_{x},%
\;q_{t_{3}}=A_{3}A_{2}^{-1}q_{t_{2}},  \tag{5.6}  \label{77}
\end{equation}%
where 
\begin{equation*}
A_{2}=\left( 
\begin{array}{ccc}
\frac{1}{2}q_{2}^{2}+\frac{1}{2}q_{3}^{2}-\alpha _{2}-\alpha _{3} & -\frac{1%
}{2}q_{1}q_{2} & -\frac{1}{2}q_{1}q_{3} \\ 
-\frac{1}{2}q_{1}q_{2} & \frac{1}{2}q_{1}^{2}+\frac{1}{2}q_{3}^{2}-\alpha
_{1}-\alpha _{3} & -\frac{1}{2}q_{2}q_{3} \\ 
-\frac{1}{2}q_{1}q_{3} & -\frac{1}{2}q_{2}q_{3} & \frac{1}{2}q_{1}^{2}+\frac{%
1}{2}q_{2}^{2}-\alpha _{1}-\alpha _{2}%
\end{array}%
\right) ,
\end{equation*}%
\begin{equation*}
A_{3}=\left( 
\begin{array}{ccc}
\begin{array}{c}
-\frac{1}{2}\alpha _{3}q_{2}^{2}-\frac{1}{2}\alpha _{2}q_{3}^{2} \\ 
+\alpha _{2}\alpha _{3}%
\end{array}
& \frac{1}{2}\alpha _{3}q_{1}q_{2} & \frac{1}{2}\alpha _{2}q_{1}q_{3} \\ 
\frac{1}{2}\alpha _{3}q_{1}q_{2} & 
\begin{array}{c}
-\frac{1}{2}\alpha _{3}q_{1}^{2}-\frac{1}{2}\alpha _{1}q_{3}^{2} \\ 
+\alpha _{1}\alpha _{3}%
\end{array}
& \frac{1}{2}\alpha _{1}q_{2}q_{3} \\ 
\frac{1}{2}\alpha _{2}q_{1}q_{3} & \frac{1}{2}\alpha _{1}q_{2}q_{3} & 
\begin{array}{c}
-\frac{1}{2}\alpha _{2}q_{1}^{2}-\frac{1}{2}\alpha _{1}q_{2}^{2} \\ 
+\alpha _{1}\alpha _{2}%
\end{array}%
\end{array}%
\right) .
\end{equation*}%
The system (\ref{77}) takes a WNSH form (\ref{40})-(\ref{42}) in the
generalizes elliptic coordinates after the following coordinate
transformation 
\begin{eqnarray*}
\rho _{1}(\lambda ) &=&-\frac{1}{2}(q_{1}^{2}+q_{2}^{2}+q_{3}^{3})+\alpha
_{1}+\alpha _{2}+\alpha _{3}, \\
\rho _{2}(\lambda ) &=&\frac{1}{2}[(\alpha _{2}+\alpha
_{3})q_{1}^{2}+(\alpha _{1}+\alpha _{3})q_{2}^{2}+(\alpha _{1}+\alpha
_{2})q_{3}^{2}]-(\alpha _{1}\alpha _{2}+\alpha _{1}\alpha _{3}+\alpha
_{2}\alpha _{3}), \\
\rho _{3}(\lambda ) &=&-\frac{1}{2}(\alpha _{2}\alpha _{3}q_{1}^{2}+\alpha
_{1}\alpha _{3}q_{2}^{2}+\alpha _{1}\alpha _{2}q_{3}^{2})+\alpha _{1}\alpha
_{2}\alpha _{3}.
\end{eqnarray*}%
Of course just the KdV case offers us infinitely many Liouville integrable
bi-Hamiltonian systems, generated from the symmetry constraint of the KdV
equation 
\begin{eqnarray}
q_{kxx}+uq_{k} &=&\alpha _{k}q_{k},\;\;k=1,...,n,  \notag \\
\frac{\delta h_{m}}{\delta u} &=&\sum_{k=1}^{n}\frac{\delta \alpha _{k}}{%
\delta u}+c=\sum_{k=1}^{n}q_{k}^{2}+c,  \TCItag{5.7}  \label{78}
\end{eqnarray}%
where $h_{m}$ is the $m$-th conserved functional of the KdV hierarchy. In
each case, we get a cofactor type hydrodynamic system.

Just to demonstrate a vast universality of this approach, our second example
is related to another soliton hierarchy, represented by the Jaulent Miodek
spectral problem \cite{j} (a special case of Antonowicz and Fordy spectral
problem \cite{fo}) 
\begin{equation}
\left( 
\begin{array}{c}
q_{i} \\ 
p_{i}%
\end{array}%
\right) _{x}=\left( 
\begin{array}{cc}
0 & 1 \\ 
\alpha _{i}^{2}-u_{1}\alpha _{i}-u_{0} & 0%
\end{array}%
\right) \left( 
\begin{array}{c}
q_{i} \\ 
p_{i}%
\end{array}%
\right) ,\;\;i=1,...,n,  \tag{5.8}  \label{79}
\end{equation}%
and its $m$-th symmetry constraint 
\begin{equation}
\frac{\delta h_{m}}{\delta u}=\frac{1}{2}\left( 
\begin{array}{c}
(q,Aq)+c \\ 
(q,q)%
\end{array}%
\right) ,  \tag{5.9}  \label{80}
\end{equation}%
Here we consider the case of $m=4$ \cite{m3} 
\begin{equation*}
h_{4}=\frac{7}{128}u_{1}^{5}+\frac{5}{16}u_{1}^{3}u_{0}-\frac{5}{32}%
u_{1x}^{2}u_{1}+\frac{3}{8}u_{0}^{2}u_{1}-\frac{1}{8}u_{1x}u_{0x}.
\end{equation*}%
By introducing the Ostrogradsky coordinates 
\begin{equation*}
q_{n+1}=u_{1},\;q_{n+2}=u_{0},
\end{equation*}%
\begin{equation*}
p_{1}=\frac{\delta H_{4}}{\delta u_{1x}}=-\frac{5}{16}u_{1}u_{1x}-\frac{1}{8}%
u_{0x},\;p_{2}=\frac{\delta H_{4}}{\delta u_{0x}}=-\frac{1}{8}u_{1x},
\end{equation*}%
Eqs. (\ref{79}), (\ref{80}) for $m=4$ are transformed into a canonical
finite dimensional Hamiltonian system with the Hamiltonian function 
\begin{eqnarray*}
H_{1} &=&\frac{1}{2}(p,p)-\frac{1}{2}(q,Aq)+\frac{1}{2}q_{n+1}(q,Aq)+\frac{1%
}{2}q_{n+2}(q,q)+cq_{n+1} \\
&&-8p_{n+1}p_{n+2}+10q_{n+1}p_{n+2}^{2}-\frac{5}{16}q_{n+1}^{3}q_{n+2}-\frac{%
3}{8}q_{n+1}q_{n+2}^{2}-\frac{7}{128}q_{n+1}^{5}.
\end{eqnarray*}%
The bi-Hamiltonian (quasi-bi-Hamiltonian) representation was found in \cite%
{m3}, \cite{y} with the second Poisson structure $\Pi _{0}$ $(\Theta _{0})$
in the form (\ref{45}) with $L(q)$ matrix of the size $(n+2)\times (n+2)$ 
\begin{equation*}
L=\left( 
\begin{array}{ccc}
A & -\frac{1}{4}q & 0_{n\times 1} \\ 
0_{1\times n} & q_{n+1} & 1 \\ 
2q^{T} & -\frac{1}{2}q_{n+2}-\frac{15}{8}q_{n+1}^{2} & -\frac{3}{2}q_{n+1}%
\end{array}%
\right) ,
\end{equation*}%
which gives us a set of $(n+2)$ component cofactor hydrodynamic systems. Let
us look at the three-component case of $n=1,$ then 
\begin{equation*}
L=\left( 
\begin{array}{ccc}
\alpha & -\frac{1}{4}q_{1} & 0 \\ 
0 & q_{2} & 1 \\ 
2q_{1} & -\frac{1}{2}q_{3}-\frac{15}{8}q_{2}^{2} & -\frac{3}{2}q_{2}%
\end{array}%
\right)
\end{equation*}%
and $cof(\lambda I-L)=I\lambda ^{2}+A_{2}\lambda +A_{3},$ where 
\begin{equation*}
A_{2}=\left( 
\begin{array}{ccc}
\frac{1}{2}q_{2} & -\frac{1}{4}q_{1} & 0 \\ 
0 & \frac{3}{2}q_{2}-\alpha & 1 \\ 
2q_{2} & -\frac{15}{8}q_{2}^{2}-\frac{1}{2}q_{3} & -q_{2}-\alpha%
\end{array}%
\right) ,
\end{equation*}%
\begin{equation*}
A_{3}=\left( 
\begin{array}{ccc}
\frac{3}{8}q_{2}^{2}+\frac{1}{2}q_{3} & -\frac{3}{8}q_{1}q_{2} & -\frac{1}{4}%
q_{1} \\ 
2q_{2} & -\frac{3}{2}\alpha q_{2} & -\alpha \\ 
-2q_{2}^{2} & \frac{15}{8}\alpha q_{2}^{2}-\frac{1}{2}q_{1}q_{2}+\frac{1}{2}%
\alpha q_{3} & \alpha q_{2}%
\end{array}%
\right)
\end{equation*}%
and the related cofactor hydrodynamic systems are 
\begin{equation}
q_{t_{2}}=A_{2}q_{x},\;q_{t_{3}}=A_{3}q_{x},%
\;q_{t_{3}}=A_{3}A_{2}^{-1}q_{t_{2}}.  \tag{5.10}  \label{81}
\end{equation}%
The system (\ref{81}) takes a WNSH form (\ref{40})-(\ref{42}) after the
following coordinate transformation\ 
\begin{eqnarray*}
q_{1} &=&[\frac{5}{2}\alpha ^{3}-5\alpha ^{2}\rho _{1}(\lambda )+\frac{3}{2}%
\alpha \rho _{1}^{2}(\lambda )-\alpha \rho _{2}(\lambda )-\rho _{3}(\lambda
)]/[2\alpha -2\rho _{1}(\lambda )], \\
q_{2} &=&2\alpha -2\rho _{1}(\lambda ), \\
q_{3} &=&-\alpha ^{2}+2\alpha \rho _{1}(\lambda )-3\rho _{1}^{1}(\lambda
)-2\rho _{2}(\lambda ).
\end{eqnarray*}

\section{Concluding remarks}

In this paper, developing the ideas of Ferapontov and Fordy \cite{f1} and
Ibort, Magri and Marmo \cite{mg2}, we have established the relation between
the bi-Hamiltonian family of St\"{a}ckel systems and the class of
hydrodynamic systems whose complete integral is constructed from a complete
solution of the related St\"{a}ckel family. Moreover, we have found the most
general admissible Riemann invariant form of such hydrodynamic systems in
separated coordinates. In a particular case of one Casimir St\"{a}ckel
family (cofactor St\"{a}ckel systems), we also presented some systematic
methods of construction of related hydrodynamic systems in arbitrary
coordinates and the explicit form of the transformation to the Riemann
invariant form. The second of the presented methods reveals a new
interesting link between soliton systems and the class of hydrodynamic
systems considered. Actually, the link is as follows: 
\begin{equation*}
soliton\;system\rightarrow St\ddot{a}ckel\;system\rightarrow
hydrodynamic\;system
\end{equation*}%
and the complete solution to the St\"{a}ckel system is a simultaneous
particular solution of both soliton and hydrodynamic system.

\bigskip

{\Large Acknowledgment}

This work was partially supported by KBN grant No 5P03B 004 20, Research
Grants Council of Hong Kong (Project No 9040466) and a grant from the City
University of Hong Kong (Project Nos 7001041, 7001178). One of the authors
(MB) also acknowledges warm hospitality at the City University of Hong Kong.

\end{document}